\documentclass[
12pt,
preprint,preprintnumbers,nofootinbib,
groupedaddress,superscriptaddress,amsmath,amssymb]{revtex4}
\usepackage{graphicx}
\usepackage{dcolumn}
\usepackage{bm}
\usepackage{amssymb}
\usepackage{amsmath}
\usepackage{epsfig}    
\usepackage{color}
\usepackage{slashed}
\usepackage{hhline}

\def\be{\begin{equation}}
\def\ee{\end{equation}}
\newcommand{\bea}{\begin{eqnarray}}
\newcommand{\eea}{\end{eqnarray}}
\newcommand{\nn}{\nonumber}

\numberwithin{equation}{section}

\begin{document}
    \rightline{KIAS-P20009, APCTP Pre2020-003}

\title{ A multi-charged particle model with local $U(1)_{\mu - \tau}$ to explain  muon $g-2$,  flavor physics, and possible collider signature}

\author{Nilanjana Kumar}
\email{nilanjana.kumar@gmail.com}
\affiliation{Department of Physics and Astrophysics, University of Delhi, Delhi 110007, India}

\author{Takaaki Nomura}
\email{nomura@kias.re.kr}
\affiliation{School of Physics, KIAS, Seoul 02455, Republic of Korea}

\author{Hiroshi Okada}
\email{hiroshi.okada@apctp.org}
\affiliation{Asia Pacific Center for Theoretical Physics (APCTP) - Headquarters San 31, Hyoja-dong,
Nam-gu, Pohang 790-784, Korea}
\affiliation{Department of Physics, Pohang University of Science and Technology, Pohang 37673, Republic of Korea}

\date{\today}

\begin{abstract}
We consider a model with multi-charged particles including vector-like fermions and a charged scalar under a local $U(1)_{\mu - \tau}$ symmetry.
We search for allowed parameter region explaining muon anomalous magnetic moment (muon $g-2$) and $b \to s \ell^+ \ell^-$ anomalies, satisfying constraints from the lepton flavor violations, $Z$ boson decays, 
meson anti-meson mixing and collider experiments.
Carrying out numerical analysis, we explore the typical size of the muon $g-2$ and Wilson coefficients to explain $b \to s \ell^+ \ell^-$ anomalies in our model when all other experimental constraints are satisfied.
We then discuss the collider physics of the multicharged vectorlike fermions, considering some benchmark points in the allowed parameter space. 
\end{abstract}
\maketitle

\newpage

\section{Introduction}
Muon anomalous magnetic moment (muon $g-2$) is analyzed with high precision, both experimentally and theoretically and 
it is a promising observable to test/confirm new physics beyond the standard model(SM).
Recently,  E989 Run 1 experiment at Fermilab(FNAL)~\cite{Abi:2021gix} provide new data of muon $g-2$ 
where the previous measurement at the E821 experiment at Brookhaven National Lab (BNL) two decades ago~\cite{Bennett:2006fi} indicates deviation from the SM prediction by $\sim 3 \sigma$.
Combining BNL result, the deviation from the SM prediction~\cite{Hagiwara:2011af,Keshavarzi:2018mgv} is given by
\begin{align}
\Delta a_\mu =(25.1\pm5.9)\times 10^{-10},
\end{align}
where the deviation reaches $4.2\sigma$ with a positive value from the SM prediction. 
Moreover, further update of  Fermilab E989  and upcoming J-PARC E34 \cite{jpark} experiment will provide the results with higher precision. 
In order to explain the deviation theoretically, several mechanisms have been proposed historically, for example, gauge contributions~\cite{Altmannshofer:2014pba, Mohlabeng:2019vrz, Abdallah:2011ew}, Yukawa contributions at one-loop level~\cite{Lindner:2016bgg}, and Barr-Zee contributions~\cite{bz} at two-loop level. 
In particular, if muon $g-2$ is related to the other phenomenologies such as neutrino mass generations, dark matter and various flavor physics, the new Yukawa interactions become important where muon $g-2$ would be explained at one-loop level through such interactions~\cite{Ma:2001mr, Okada:2013iba, Baek:2014awa, Okada:2014nsa, Okada:2014qsa, Okada:2015hia, Okada:2016rav, Nomura:2016rjf, Ko:2016sxg, Baek:2016kud, Nomura:2016ask, Lee:2017ekw, Chiang:2017tai, Das:2017ski, Nomura:2017ezy, Nomura:2017tzj, Cheung:2017kxb, Cheung:2018itc, Cai:2017jrq, Chakrabarty:2018qtt, CarcamoHernandez:2019xkb,Chen:2019nud,Nomura:2017ohi,Baumholzer:2018sfb, Nomura:2019btk, Chen:2020ptg,Lindner:2016bgg,Barman:2018jhz,Arnan:2019uhr,Calibbi:2019bay,Darme:2020hpo,Altmannshofer:2016oaq,Li:2019xmi} 
(see also recent approaches after new result of FNAL~\cite{Arcadi:2021cwg,Zhu:2021vlz,Han:2021gfu,Baum:2021qzx,Bai:2021bau,Das:2021zea,Lu:2021vcp,Ge:2021cjz,Brdar:2021pla,Buen-Abad:2021fwq,Zu:2021odn,Amaral:2021rzw,Endo:2021zal,Ahmed:2021htr,Abdughani:2021pdc,VanBeekveld:2021tgn,Cox:2021gqq,Wang:2021bcx,Gu:2021mjd,Cao:2021tuh,Yin:2021mls,Han:2021ify,Aboubrahim:2021xfi,Yang:2021duj,Chakraborti:2021bmv,Ferreira:2021gke,Wang:2021fkn,Li:2021poy,Cadeddu:2021dqx,Calibbi:2021qto,Chen:2021vzk,Escribano:2021css,Eung:2021bef,Arcadi:2021yyr,Chen:2021jok,Nomura:2021oeu}).  
In such case, one has to simultaneously satisfy several constraints of lepton flavor violations (LFVs), such as $\ell_i\to \ell_j\gamma$, $\ell_i\to\ell_j\ell_k\bar\ell_\ell$ ($i,j,k,\ell=(e,\mu,\tau)$), and lepton flavor conserving(violating) $Z$ boson decays $Z\to\ell\bar\ell'$, $Z\to\nu\bar\nu'$~\cite{pdg}. 
In particular, $\ell_\mu\to \ell_e\gamma$ process give the most stringent constraint where the current upper bound on the branching ratio is $4.2\times10^{-13}$~\cite{TheMEG:2016wtm}, and its future bound will reach the sensitivity at $6\times10^{-14}$~\cite{Renga:2018fpd}. In addition, $Z$ boson decays will be tested by future experiments such as CEPC~\cite{cepc}.
Previously, we analyzed models introducing multi-charged fields (scalars and vector-like leptons) with general $U(1)_Y$ hypercharges to obtain  positive muon $g-2$ 
and explored the parameter region satisfying several experimental constraints~\cite{Nomura:2019btk}. Another interesting study includes 
a new $U(1)'$ in order to explain the same \cite{Kawamura:2019rth}.

{
Another interesting hints of new physics are
experimental anomalies of semileptonic $B$-meson decay; deviations in the measurements of the angular observable $P'_5$ in the decay of the $B$ meson 
($B \to K^* \mu^+ \mu^-$)~\cite{DescotesGenon:2012zf, Aaij:2015oid, Aaij:2013qta,Abdesselam:2016llu, Wehle:2016yoi}, the ratio of branching fractions, $R_K = BR(B^+ \to K^+ \mu^+\mu^-)/ BR(B^+ \to K^+ e^+e^-)$~\cite{Hiller:2003js, Bobeth:2007dw, Aaij:2014ora}, 
and $R_{K^*} = BR (B \to K^* \mu^+ \mu^-)/ BR (B \to K^* e^+ e^-)$~\cite{Aaij:2017vbb}. 
Various global fits to corresponding Wilson coefficients are also carried out~\cite{Ciuchini:2019usw,Descotes-Genon:2015uva,Alguero:2019ptt,Aebischer:2019mlg}, 
indicating that negative contribution to Wilson coefficient associated with $(\bar s_R \gamma^\mu b_L)( \bar \mu \gamma_\mu \mu)$ operator is preferred in explaining the anomalies.
We can explain the anomalies by introducing $U(1)_{\mu - \tau}$ gauge symmetry when we include 
some extra field contents such as vector-like quarks~\cite{Altmannshofer:2014cfa,Altmannshofer:2015mqa,Altmannshofer:2016jzy,Ko:2017yrd,Chen:2017usq, Arcadi:2018tly,Hutauruk:2019crc,Crivellin:2015mga}.

Hence, it is worthwhile to consider a model with multi-charged particles-- vector-like quarks, vector-like leptons and charged scalar fields-- under local $U(1)_{\mu - \tau}$ framework
where we can combine the ideas  in the model discussed in ref.~\cite{Nomura:2019btk} and ref.~\cite{Ko:2017yrd,Hutauruk:2019crc}.
Advantages of this scenario are as follows: (1) we can constrain the flavor structure of Yukawa couplings associated with extra fermions in order to suppress the constraints from lepton flavor violations (LFVs), 
(2) we have more contributions to muon $g-2$ from one loop diagrams with $Z'$ and vector-like leptons, 
(3) collider signature is controlled by $U(1)_{\mu - \tau}$ charge assignment to give predictions.
We then investigate if both muon $g-2$ and $B$-anomalies can be 
explained simultaneously by analyzing the correlation among the parameters taking into account experimental constraints, and discuss collider physics to show possible signatures of this scenario.

In this paper, we discuss the model introducing multi-charged fields (scalars and fermions) under local $U(1)_{\mu - \tau}$ framework, 
as an extension of the model in ref.~\cite{Nomura:2019btk} and in ref.~\cite{Ko:2017yrd,Hutauruk:2019crc}.
We investigate contributions to muon $g-2$ from one-loop diagrams including the new particles such as vector-like lepton, charged scalar and $Z'$ boson. 
Extra vector-like quarks are introduced and Wilson coefficient is calculated to explain $B$-anomalies.
Constraints from meson anti-meson mixing are discussed in addition to LFV and $Z$ decays.
Then we explore the parameter region accommodating both muon $g-2$ and $B$-anomalies.
We search for the parameters satisfying all the constraints and { from the allowed model parameters
we consider the benchmark points (BP's) for the collider study.}
}
 
Since the multi-charged fields can be produced at the Large Hadron Collider (LHC),
the signature of the exotic charged particles are also explored. We particularly focus on the 
LHC signatures of exotic lepton doublet. Here the exotic leptons decay $via$ the charged scalar, which 
in turn produces different collider signatures w.r.t the standard scenario, where exotic 
leptons (singly charged) decay into SM particles directly ($W\nu$, $Z\ell$ and $H\ell$)~\cite{Sirunyan:2019ofn}. 
We will show that a small mass difference between the charged scalar and the exotic lepton is naturally
{ favored by the sizable muon ($g-2$). }
Hence, the collider signature of this particular model will contain very soft muons. We particularly focus on 
the signature of two oppositely charged muon and tau pair at LHC.

This paper is organized as follows.
In Sec.~II, we show setup of the model and formulate the Wilson coefficient for $B$-decay, meson anti-meson mixing, LFV's, muon $g-2$ and $Z$ boson decays.
In Sec.~III, we perform numerical analysis searching for the allowed region of parameter space.
In Sec.~IV, we discuss possible extension of the model introducing $U(1)_{\mu - \tau}$ gauge symmetry and discuss collider physics signature.
We conclude in Sec.~V.
\section{Model setup and formalism}
\begin{table}[t]
\begin{tabular}{|c|c|c|c|c|c|c|c|c||c|c|c|}
\hline\hline  
 &~ $L_{L_\mu}$ ~&~ $L_{L_\tau}$ ~&~ $e_{R_\mu}$ ~&~ $e_{R_\tau}$ ~&~ $\nu_{R_\mu}$ ~&~ $\nu_{R_\tau}$ ~&~$L'_{}$~&~$Q'_{}$ ~& ~$H$~ & ~$s^{+}$ & ~$\varphi$~\\\hline 
$SU(3)$ & $\bm{1}$ &  $\bm{1}$& $\bm{1}$& $\bm{1}$& $\bm{1}$& $\bm{1}$& $\bm{1}$& $\bm{3}$   & $\bm{1}$ & $\bm{1}$ & $\bm{1}$   \\\hline 
$SU(2)_L$ & $\bm{2}$ &  $\bm{2}$& $\bm{1}$& $\bm{1}$ & $\bm{1}$& $\bm{1}$ & $\bm{2}$& $\bm{2}$   & $\bm{2}$ & $\bm{1}$ & $\bm{1}$  \\\hline 
$U(1)_Y$   & $-\frac12$ & $-\frac12$ & $-1$ & $-1$ & $0$ & $0$ & $-\frac{3}2$& $-\frac{5}6$ & $\frac12$ & $+1$  & $0$   \\\hline
$U(1)_{\mu -\tau}$ & $1$ & $-1$ & $1$ & $-1$ & $1$ & $-1$ & $1+x$ & $x$ & 0 & $-x$ & $y$ \\ \hline
\end{tabular}
\caption{Charge assignments of fields under $SU(2)_L\times U(1)_Y \times U(1)_{\mu - \tau}$ for the extended model. 
We introduce three generations of vector-like fermions $L'$ and $Q'$.  }
\label{tab:1}
\end{table}

We consider a model with gauge symmetry $G_{\rm SM} \times U(1)_{\mu - \tau}$ where $G_{\rm SM}$ is the SM gauge symmetry 
and $U(1)_{\mu - \tau}$ is an extra gauge symmetry.
In our set up of the model, we introduce isospin doublet fermions $L'_a\equiv[\psi^{-}_a,\psi^{--}_a]^T (a=1-3)$,
 $Q'_a\equiv[q'^{-1/3}_a,q'^{-4/3}_a]^T\equiv [u'_a,d'_a]^T$ and a singly-charged boson $s^{+}$ as shown
in Table~\ref{tab:1}~\footnote{ We introduce three generations of vector like fermions just to match with the number of generations for SM fermions. In principle, we can explain anomalies discussed in the paper by one generation of vector like fermion.}; here $x$ and $y$ for $U(1)_{\mu - \tau}$ are any real number and the SM quarks are not charged under $U(1)_{\mu -\tau}$.
 For vector-like fermions, we introduce three generations to match with the SM.
 We also introduce three right-handed neutrinos with $U(1)_{\mu - \tau}$ charge~\footnote{In this paper we do not discuss neutrino mass. Neutrino masses under $U(1)_{\mu - \tau}$ can be found e.g. in ref.~\cite{Asai:2017ryy,Asai:2018ocx}. }
Here, we also introduced scalar field $\varphi$ with non-zero VEV to break $U(1)_{\mu -\tau}$ spontaneously. 
The Lagrangian involving the interaction of new particles and SM and the potential is given by,
\begin{align}
-\mathcal{L}^n_{Y}  &= f_{2a} \bar L_{L_2} L'_{R_a} s^+ + g_{ia} \bar Q_{L_i} Q'_{R_a} s^+ 
+h_{ij} \bar L^c_{L_i}\cdot  L_{L_j} s^+ +  k_{ij} \overline{\nu^c_{R_i}} e_{R_j}  s^+ \nonumber \\
& + M_{Q'_a} \bar Q'_{L_a} Q'_{R_a}
+ M_{\psi_a} \bar L'_{L_a} L'_{R_a}
+ {\rm h.c.}\nn\\
&=f_{2a}[\bar \nu_{2} P_R\psi^{-}_a s^+ + \bar\ell_2 P_R\psi^{--}_a s^+]
+g_{ia}[\bar u_{i} P_R u'_a s^- + \bar d_i P_R d'_a s^+]  + h_{ij}[\bar \nu^c_{i} P_L \ell_j s^+ - \bar \ell^c_i P_L \nu_j s^+]  \nn\\
& + k_{ij} \overline{\nu^c_{R_i}} e_{L_j} s^+ + M_{Q'_a} \bar Q'_{L_a} Q'_{R_a}
+ M_{\psi_a} \bar L'_{L_a} L'_{R_a}+ {\rm h.c.}
, \label{Eq:lag-yukawa}  \\ 
{\cal V}&= \mu_H^2|H|^2 + \mu_S^2 |s^+|^2
+\lambda_H|H|^4 +\lambda_s|s^+|^4   +\lambda_{Hs}|H|^2|s^+|^2 + \mu^2_\varphi |\varphi|^2 + \lambda_{\varphi} |\varphi|^4 \nn \\
& \quad  + \lambda_{H \varphi} |H|^2 |\varphi|^2 + \lambda_{S \varphi} |s^+|^2 |\varphi|^2,\label{eq:pot}
\end{align}
where $(i,j,a)=1-3$ are generation indices, ($\cdot\equiv i\sigma_2$), $\sigma_2$ being the second Pauli matrix, and $L'_{L[R]_a}(Q'_{L[R]_a}) \equiv P_{L[R]} L'_a(Q'_a)$.
The SM Yukawa term $y_{\ell_{ii}}\bar L_{L_i}e_{R_i}H$
provides masses for the charged leptons {$(m_{\ell_i}\equiv y_{\ell_{ii}}v/\sqrt2$)} by developing a nonzero 
vacuum expectation value (VEV) of $H$, which is denoted by $\langle H\rangle\equiv v/\sqrt2$.
The exotic fermion mass eigenvalues are respectively $M_{Q'}, M_\psi$ for $Q', L'$.
 We expect that the interaction term involving $h_{ij}$ plays a role in $s^+$ decay into the SM fields appropriately. 
 However, since this term gives negative contribution to the muon $g-2$, we assume the scale of $h_{ij}$ is not so large. It implies that we do not discuss LFVs and muon $g-2$ of this term.
 Note also that non-zero components of $h_{ij}$ and $k_{ij}$ are changed by our choice of parameter $x$, and thus decay pattern of $s^+$ depends on $x$.
More concretely, structure of the third and fourth terms in Eq.~\eqref{Eq:lag-yukawa} depends on value of $x$ such that 
\begin{align}
& h_{ij} \bar L^c_{L_i}\cdot  L_{L_j} s^+ =  h_{\{12, 13 \}} \bar L^c_{L_{\{e,e\} } } \cdot L_{L_{\{\mu,\tau\}}} s^+ \quad \text{for $x = \{1, -1 \}$}, \\
& k_{ij} \overline{\nu^c_{R_i}}  e_{R_j} s^+ =  k_{ \{ 12(21), 13(31), 22, 33 \}}    \overline{ \nu^c_{R_{\{ e(\mu), e(\tau), \mu, \tau \}}} } e_{R_{\{ \mu(e), \tau(e), \mu, \tau \} }} s^+ \quad \text{for $x = \{1, -1, 2,  -2 \}$},
\end{align}
where we cannot have the Yukawa interaction for other $x \neq 0$. 
Thus, the decay pattern of $s^+$ is determined by the choice of $x$ where we consider our right-handed neutrinos are assumed to be light so that $s^+$ can decay into states containing them.
For $x = -2$, constraint from collider experiment is weaker since $s^\pm$ only decays into third generation of leptons 
while we have stronger constraint for $x = \pm 1$ or $2$, since it decays into electron and/or muon.
Therefore we chose $x = -2$ in our numerical analysis.
 In addition we do not have extra term in any choice of $y$, where $x\neq0$ and $y\neq0$.

In scalar sector, we assume coupling $\lambda_{H \varphi}$ is small so that mixing between $\varphi$ and $H$ is negligible for simplicity.
Under the assumption, the VEV of $\varphi$ is simply given by $v_\varphi \simeq \sqrt{-\mu_\varphi^2/\lambda_\varphi}$.
After $\varphi$ developing a VEV, we have massive $Z'$ boson whose mass is given by
\begin{equation}
m_{Z'} = y g' v_\varphi,
\end{equation}
where $g'$ is gauge coupling associated with $U(1)_{\mu - \tau}$.
The mass eigenvalue of $s^+$ is given by 
\begin{align}
m_S = \mu_S^2+ \frac{\lambda_{Hs}}{2}  v^2+ \frac{\lambda_{S \varphi }}{2}  v_\varphi^2.
\end{align}
In our numerical analysis we take $m_S$ as a free parameter.

\subsection{{\boldmath$M-\overline M$} mixing}
The parameter space of our model get constrained from the neutral meson mixings, where the 
VLQ's appear in the loop. The relevant expressions   
as shown in~\cite{Gabbiani:1996hi}, are
\begin{align}
&\Delta M_Q\approx  \frac{ m_Q f_Q^2}{3(4\pi)^2}
\sum_{a,b=1}^3{{\rm Re}[g_{ka} g_{ai}^* g_{jb} g_{b \ell }^*]} F_{\rm box}(M_{Q'_a},M_{Q'_b},m_s),\\
& F_{\rm box}(m_1,m_2,m_3)  =
\int [dx]^3 \frac{z}{x m_1^2+y m_2^2+z m^2_3},
\end{align}
where $\int [dx]^3 \equiv \int_0^1 dx dy dz \delta(1-x-y-z)$,  $B_s-\bar B_s$ mixing corresponds to $(i,j,k,\ell)=(2,3,3,2)$, $B_d-\bar B_d$ 
mixing corresponds to $(i,j,k,\ell)=(1,3,3,1)$,
$K-\bar K$ and $D-\bar D$ to $(i,j,k,\ell)=(1,2,2,1)$.
{The neutral meson mixing {formulas} should be lower than the experimental bounds as given in
~\cite{Gabbiani:1996hi, Agashe:2014kda}:
\begin{align}
\Delta m_K &  \lesssim 3.48\times10^{-15} \ [{\rm GeV}], \label{eq:kk}\\
3.29\times10^{-13} \ [{\rm GeV}]&\lesssim\Delta m_{B_d} +\Delta m_{B_d}^{SM}  \lesssim 3.37\times10^{-13} \ [{\rm GeV}],\\
1.16\times10^{-11} \ [{\rm GeV}] &\lesssim\Delta m_{B_s} + \Delta m_{B_s}^{SM}   \lesssim 1.17\times10^{-11} \ [{\rm GeV}],\\
\Delta m_D & \lesssim 6.25\times10^{-15} \ [{\rm GeV}],\label{eq:dd} 
\end{align}
where we have taken 3$\sigma$ interval and $m_{M}$ and $f_M$ are the meson mass and the meson decay constant, respectively.
The following values of the parameters are used in our analysis: $f_K \approx 0.156$ GeV,
$f_{B_d(B_s)} \approx 0.191(0.274)$ GeV~\cite{DiLuzio:2017fdq, DiLuzio:2018wch},
  $f_{D} \approx 0.212$ GeV,
$m_K \approx 0.498$ GeV, $m_{B_d(B_s)} \approx 5.280(5.367)$ GeV, and
$m_{D} \approx 1.865$ GeV.
%
The SM contributions are given by~\cite{Amhis:2016xyh}: 
\begin{align}
2.96\times10^{-13} \ [{\rm GeV}]&\lesssim\Delta m_{B_d}^{SM}  \lesssim 5.13\times10^{-13} \ [{\rm GeV}],\\
1.06\times10^{-11} \ [{\rm GeV}] &\lesssim\Delta m_{B_s}^{SM}  \lesssim 1.44\times10^{-11} \ [{\rm GeV}].
\end{align}
Subtracting the SM contributions from the experimental results,
and one finds the following bounds:
\begin{align}
-1.85\times10^{-13} \ [{\rm GeV}]&\lesssim\Delta m_{B_d}  \lesssim 4.05\times10^{-14} \ [{\rm GeV}],\\
-2.77\times10^{-12} \ [{\rm GeV}] &\lesssim\Delta m_{B_s}  \lesssim 1.07\times10^{-12} \ [{\rm GeV}].
\end{align}
\subsection{{\boldmath$b\to s\ell_i\bar\ell_j$} decay}

\begin{figure}[t]
\centering
\includegraphics[width=120mm]{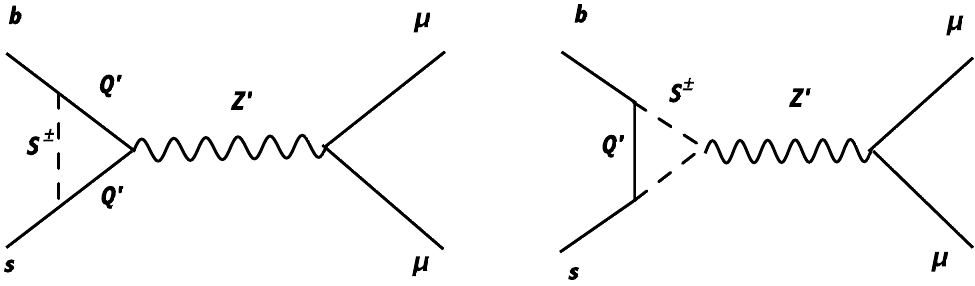} 
\caption{Diagrams that contributes to $\Delta C_9^\mu$.} 
\label{fig:C9Zp}
\end{figure}

In our model we apply the same mechanism in refs.~\cite{Ko:2017yrd, Hutauruk:2019crc} to generate $\Delta C_9^\mu$ using $Z'$ interaction; 
other mechanisms with $U(1)_{\mu -\tau}$ can be found in e.g. refs~\cite{Crivellin:2015mga,Altmannshofer:2014cfa}.
We obtain contribution to $\Delta C_9^{\mu}$ from diagrams in Fig.~\ref{fig:C9Zp}. 
Then we obtain the contribution to $\Delta C_9^{\mu, Z'}$ as in Ref~\cite{Ko:2017yrd, Hutauruk:2019crc} 
\begin{align}
\Delta C_9^{\mu, Z'} & \simeq \frac{x g'^2}{(4\pi)^2 m_{Z'}^2 C_{\rm SM}}
\sum_{a=1}^{3} g_{3a} g^*_{a2}
\int[dx]^2 \ln \left(\frac{\Delta[M_{Q'_a},m_S]}{\Delta[m_S,M_{Q'_a}]}  + \frac{M^2_{Q'_a}}{x m_S^2 + y M^2_{Q'_a}} \right), \nn \\
C_{\rm SM} &\equiv \frac{V_{tb} V^*_{ts} G_{\rm F} \alpha_{\rm em}}{\sqrt2 \pi},  \nn \\
\Delta[m_1,m_2] &= x m_1^2 + y m_{2}^2, 
\label{eq:c9-formula}
\end{align}
where $\int [dx]^2 \equiv \int_0^1 dx dy \delta(1-x-y)$ and quark masses are ignored.
We can obtain $\Delta C_9^{\mu, Z'} \sim -1$ satisfying all the experimental constraints as shown in refs.~\cite{Ko:2017yrd, Hutauruk:2019crc} 
with $M_{Q'_{a}} = \mathcal{O}(1)$ TeV, $m_S = \mathcal{O}(100)$ GeV and $m_{Z'} = \mathcal{O}(100)$ GeV, where
$Z'$ contribution to muon $g-2$ is small in this region.

{ Here, we simplify the above formula by carrying out integration 
\begin{align}
\Delta C_9^{\mu, Z'} & \simeq \sum_{a} \frac{g_{3a} g^*_{a2} x g'^2}{ 2 (4\pi)^2 m_{Z'}^2 C_{\rm SM}}
 \label{eq:c9-approx}
\end{align}
}

In addition we also obtain Effective Lagrangian to induce $b\to s\ell\bar\ell$ decay $via$ box diagram such that
\begin{align}
{\cal L}^{\rm [box]}=- \sum_{a,b} \frac{g_{2a}g^*_{a3} f_{2b} f^*_{b2}}{4(4\pi)^2} (\bar s\gamma_\mu P_L b)(\bar\ell_2 \gamma^\mu \ell_2 - \bar\ell_2 \gamma^\mu \gamma_5\ell_2) F_{\rm box}(M_{Q'_a} , M_{\psi_b} ,m_s),
\end{align}
which corresponds to ${\cal O}_9=-{\cal O}_{10}$~\cite{Descotes-Genon:2015uva}.
\begin{align}
\Delta C_9^{\mu {\rm [box]}}=- \Delta C_{10 }^{\mu {\rm [box]}}\approx \sum_{a,b} \frac{g_{2a} g^*_{a3} f_{2 b} f^*_{b 2}}{4 (4\pi)^2 C_{SM}}F_{\rm box}(M_{Q'_a} , M_{\psi_b} ,m_s),
\end{align}
where $C_{SM}\equiv \frac{V_{tb} V^*_{ts} G_F \alpha_{em}}{\sqrt{2} \pi}$.
In total, we obtain new physics contribution to the Wilson coefficient, $\Delta C_9^{\ell}$, as 
\begin{equation}
\Delta C_9^{\mu} = \Delta C_9^{\mu, Z'} + \Delta C_9^{\mu {\rm [box]}}.
\end{equation}
Furthermore we should take into account the diagrams replacing $Z'$ by $Z$ in Fig.~\ref{fig:C9Zp} which induce flavor universal contributions to $C_9$ and $C_{10}$ $via$ Z boson exchange.
Calculating the diagrams we obtain
\begin{align}
& \Delta C_9 (Z) \simeq  \sum_{a} \frac{g_{3a} g^*_{a2}  g_2^2}{ 4 (4\pi)^2 m_{Z}^2 c_W^2 C_{\rm SM}} \left( -\frac12 +\frac43 s_W^2 \right) \left( - \frac12 + 2 s^2_W \right), \\
& \Delta C_{10} (Z) \simeq \sum_{a} \frac{g_{3a} g^*_{a2}  g_2^2}{ 8 (4\pi)^2 m_{Z}^2 c_W^2 C_{\rm SM}} \left( -\frac12 +\frac43 s_W^2 \right),
\end{align}
where $c_W = \cos \theta_W$ with $\theta_W$ being Weinberg angle.
Since structures of $C_{9,10}(Z)$ are similar to $\Delta C_9^{\mu, Z'}$ we obtain the relation
\begin{align}
& \frac{\Delta C_9 (Z)}{\Delta C_9^{\mu, Z'}} \simeq \frac{g_2^2}{m_Z^2 c_W^2} \frac{m_{Z'}^2}{x g'^2} \frac12 \left( -\frac12 +\frac43 s_W^2 \right) \left( - \frac12 + 2 s^2_W \right), \\
& \frac{\Delta C_{10} (Z)}{\Delta C_9^{\mu, Z'}} \simeq \frac{g_2^2}{m_Z^2 c_W^2} \frac{m_{Z'}^2}{x g'^2} \frac14 \left( -\frac12 +\frac43 s_W^2 \right). 
\end{align}
Then, the $b\to s \mu \bar \mu$ anomalies can be explained by $\Delta C_9^{\mu, Z'} = -0.97$ as the best fit value, $[-1.12,-0.81]$ at 1$\sigma$, and $[-1.27,-0.65]$ at 2$\sigma$ interval~\cite{Aebischer:2019mlg}.
The flavor universal $\Delta C_9 (Z)$ is much smaller than $\Delta C_9^{\mu, Z'}$ due to the suppression factor $(-1/2 + 2 s_W^2)$. 
For $\Delta C_{10}(Z)$, we consider constraint from $B_s \to \mu^+ \mu^-$ measurement. 
Recent LHCb measurement of the branching ratio is given by~\cite{LHCb:2021vsc,LHCb:2021awg}
\begin{equation}
\label{eq:BRBsExp}
BR(B_s^0 \to \mu^+ \mu^-)^{\rm exp} = (3.09^{+0.46 + 0.15}_{-0.43-0.11}) \times 10^{-9},
\end{equation}
where first uncertainty is statical and the second one is systematic.
We can estimate the branching ratio in the model such that~\cite{Hiller:2014yaa}
\begin{equation}
BR(B_s^0 \to \mu^+ \mu^-)^{\rm th} = |1-0.24 \Delta C_{10}^{\mu \mu}|^2 BR(B_s^0 \to \mu^+ \mu^-)^{\rm SM},
\end{equation} 
where $BR(B_s^0 \to \mu^+ \mu^-)^{\rm SM} = (3.65 \pm 0.23) \times 10^{-9}$ is the theoretical predication in the SM~\cite{Bobeth:2013uxa}.
In the numerical analysis we impose that the branching ratio in our model is within $1\sigma$ region in Eq.~\eqref{eq:BRBsExp}.
Note also that $x<0$ is preferred since we realize positive $C_{10}$ to fit the data.

\subsection{Lepton flavor violations and muon anomalous magnetic moment}
\label{lfv-lu}


In our model we do not have lepton flavor violation from Yukawa coupling $f_{ia}$ since only components associate with muon, $f_{2a}$, are non-zero.
We thus only focus on the contribution to muon $g-2$ from the Yukawa interactions. \\
 %
 
\noindent
{\it \bf The muon anomalous magnetic moment} ($\Delta a_\mu$): 
We can estimate scalar loop contribution to the muon anomalous magnetic moment through(muon $g-2$), which is given by 
\begin{align}
\Delta a^{S}_\mu \approx -m_\mu (a_L+a_R)_{22}.\label{eq:damu}
\end{align}
The amplitude $a_{L/R}$ can be expressed as,
\begin{align}
 (a_{L})_{22} \approx (a_{R})_{22}
\approx
- m_{\mu} \sum_{a=1-3} \frac{f_{2a} f^*_{a2}}{(4\pi)^2}  
{\left[ F(M_{\psi^{--}_a} , m_{S}) + 2 F(m_{S},M_{\psi^{--}_a})\right]}, 
\label{eq:lfv-lp}
\end{align} 
where $M_{\psi^{--}}\equiv M_\psi$.

{It is worthwhile considering $\Delta a_\mu$ $via$ $Z'$, even though it would not definitely be needed because we already have the contribution $via$ $f_{2a}$ and preferred mass range is lighter than that for the $B$ anomalies.}  
The $Z'$ boson loop contribution is obtained as~\cite{Nomura:2018cle} 
\begin{align}
\Delta a^{Z'}_\mu=\frac{g'^2m_\mu^2}{4\pi^2}\int_0^1  dx \frac{x^2(1-x)}{x^2 m^2_\mu +(1-x) m_{Z'}^2}.
\end{align}
In total muon $g-2$ is given by
\begin{equation}
\Delta a_\mu = \Delta a^{S}_\mu +  \Delta a^{Z'}_\mu.
\end{equation}
The measured value show $3.3\sigma$ deviation from the SM prediction, given by
$\Delta a_\mu=(26.1\pm8)\times10^{-10}$~\cite{Hagiwara:2011af}, which is also 
a positive value. {Note here that} the charged scalar contribution through $h_{23}$ is negligible 
as we consider $h_{23}$ to be small, as discussed below {Eq.(\ref{eq:pot})}. 
\begin{figure}[tb]
\begin{center}
\includegraphics[width=120mm]{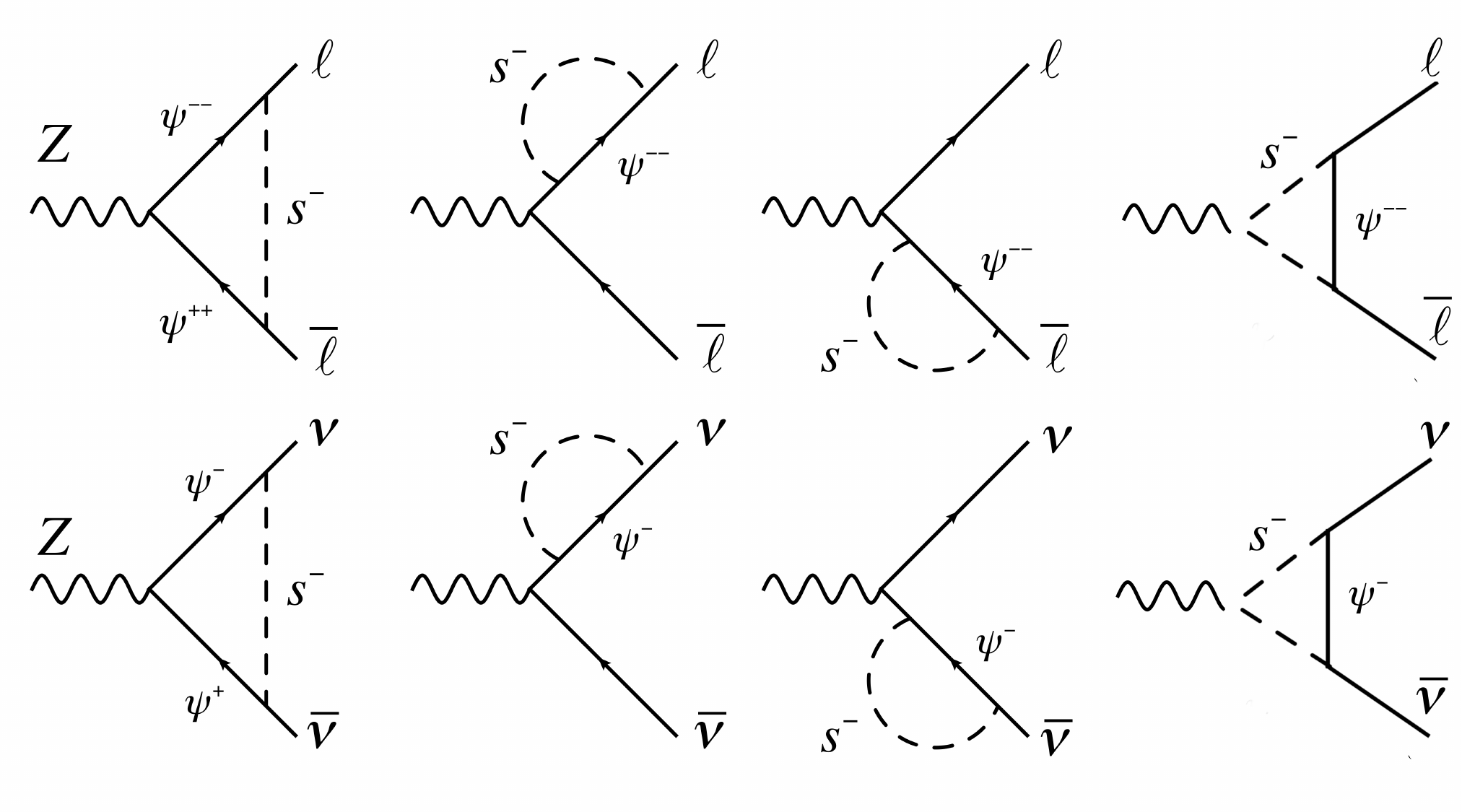}
\caption{Feynman diagrams for $Z\to \ell_i\bar\ell_j$ (up) and $Z \to \nu_i \bar \nu_j$ (down).}
\label{fig:zto2ell}
\end{center}
\end{figure}
\subsection{Flavor-{Conserving} Leptonic $Z$ Boson Decays}\label{subsec:Zll}
Here, we consider the $Z$ boson decay into two leptons through the Yukawa terms involving $f_{2a}$ at one-loop level~\cite{Chiang:2017tai}.
Since some components of $f_{2a}$ are expected to be large in order to obtain 
the sizable $\Delta a_\mu$, the experimental bounds on
$Z$ boson decays could be of concern at one loop level. 
Note that $Z$ boson decays are modified only when second generation of leptons are involved due to $U(1)_{\mu -\tau}$ symmetry.
{This is why we consider the flavor conserving processes of $Z$ boson only.}

First of all, the relevant Lagrangian is given by~\footnote {We neglect one-loop contributions in the SM.}
\begin{align}
{\cal L}&\sim
\frac{g_2}{c_w} \left[\bar\ell\gamma^\mu \left(-\frac12 P_L+s_W^2\right)\ell
+\frac12\bar\nu\gamma^\mu P_L\nu
 \right] Z_\mu\nn\\
&+ \frac{g_2}{c_w} \left[
 \left(-\frac12 P_L+ s_W^2\right)\bar\psi^{+}\gamma^\mu \psi^{-}
+ \left(-\frac12 P_L+ 2 s_W^2\right)\bar\psi^{++}\gamma^\mu \psi^{--}
 \right] Z_\mu\nn\\
 & +i\frac{g_2  s_W^2}{c_W}(s^+\partial^\mu s^{-} - s^{-}\partial^\mu s^{+})Z_\mu,
\end{align}
where $s(c)_W\equiv\sin(\cos)\theta_W\sim0.23$ 
stands for the sine (cosine) of the Weinberg angle. 
The decay rate of the SM at tree level is then given by
\begin{align}
&{\rm \Gamma}(Z\to\ell^-_i\ell^+_j)_{SM}
\approx
\frac{m_Z}{12\pi} \frac{g_2^2}{c_W^2}  \left(s_W^4-\frac{s_W^2}2 + \frac18 \right)\delta_{ij},\\
&{\rm \Gamma}(Z\to\nu_i\bar\nu_j)_{SM}
\approx
\frac{m_Z}{96\pi} \frac{g_2^2}{c_W^2} \delta_{ij}.
\end{align}
Combining all the diagrams in Fig.~\ref{fig:zto2ell},
the ultraviolet divergence cancels out and only the finite part 
remains~\cite{Chiang:2017tai} and is given by,
\begin{align}
&\Delta{\rm \Gamma}(Z\to\mu^- \mu^+)
\approx
\frac{m_Z}{12\pi} \frac{g_2^2}{c_W^2}
\left[ 
\frac{|B_{22}^{\ell}|^2}{2} - {\rm Re}[A_{22} (B^{\ell})^*_{22}]
-\left(
-\frac{s_W^2}2 + \frac18\right) \right] \label{eq:Zll},\\
&\Delta{\rm \Gamma}(Z\to\nu_\mu \bar\nu_\mu)
\approx
\frac{m_Z}{24\pi} \frac{g_2^2}{c_W^2}
\left[ 
{|B_{22}^{\nu}|^2} 
-\frac{1}{4}\right] \label{eq:Znunu},
\end{align}
where,
\begin{align}
&A_{22}\approx s^2_W,
\quad B^{\ell}_{22}\approx \frac{1}{2} - \sum_{a} \frac{f_{2a} f^\dag_{a2}}{(4\pi)^2} G^{\ell}(M_{\psi_a},m_S),
\quad B^{\nu}_{22}\approx \frac{1}{2} + \sum_{a} \frac{f_{2a} f^\dag_{a2}}{(4\pi)^2} G^{\nu}(M_{\psi_a},m_S),
\\
&G^{\ell}(M_{\psi_a},m_S)\approx- s^2_W\left(-\frac12 +s_w^2\right) H_1(M_{\psi_a},m_S)
-\left(-\frac12 +s_w^2\right)^2 H_2(m_{\psi_a},m_S) \nonumber \\
& \qquad \qquad  \qquad+\left(-\frac12 +2 s_w^2\right) H_3(M_{\psi_a},m_S),\\
&G^{\nu}(M_{\psi_a},m_S)\approx-s^2_W\left(-\frac12 +s_w^2\right) H_1(M_{\psi_a},m_S)
- \frac12 H_2(M_{\psi_a},m_S)
+\left(-\frac12 + s_w^2\right) H_3(M_{\psi_a},m_S),
\\
&H_1(m_1,m_2)= \frac{m_1^4-m_2^4+4 m_1^2 m_2^2\ln\left[\frac{m_2}{m_1}\right]}{2(m_1^2-m_2^2)^2},\\
&H_2(m_1,m_2)= \frac{m_2^4 - 4m_1^2 m_2^2 +3m_1^4 - 4 m_2^2(m_2^2-2m_1^2)\ln[m_2]-4m_1^4\ln[m_1]}{4(m_1^2-m_2^2)^2},\\
&H_3(m_1,m_2)=m_1^2\left( \frac{m_1^2-m_2^2 + 2 m_2^2\ln\left[\frac{m_2}{m_1}\right]}{(m_1^2-m_2^2)^2}\right).
\end{align}
Notice here that the upper indices of $B$ and $G$;  $\ell,\nu$, respectively represent pairs of the
{muon and muon-}neutrino final states.
We consider $\psi$ as  $\psi^{--}$ inside the argument of $G^\ell$, while $\psi$ as $\psi^{-}$ inside the argument of $G^\nu$.
The current bounds on the lepton-flavor-(conserving)changing $Z$ boson decay 
branching ratios at 95 \% CL are given by \cite{pdg}:
\begin{align}
& \Delta {\rm BR}(Z\to {\rm Invisible})\approx  \sum_{i,j=1-3}\Delta {\rm BR}(Z\to\nu_i\bar\nu_j)< \pm5.5\times10^{-4} ,
\label{eq:zmt-con}\\
& \Delta   {\rm BR}(Z\to \mu^\pm\mu^\mp) <  \pm6.6\times10^{-5} ~,
\end{align}
where $\Delta {\rm BR}(Z\to f_i\bar f_j)$ ($i= j$) is defined by
\begin{align}
\Delta {\rm BR}(Z\to f_i \bar f_j)\approx 
\frac{{\rm \Gamma}(Z\to f_i \bar f_j)- {\rm \Gamma}(Z\to f_i \bar f_j)_{SM}}
{\Gamma_{Z}^{\rm tot}},
\end{align}
where the total $Z$ decay width $\Gamma_{Z}^{\rm tot} = 2.4952 \pm 0.0023$~GeV~\cite{pdg}.
We consider all these constraints in the numerical analysis in the next section.

{
\subsection{Constraints for $Z'$ interaction}
Here we discuss experimental constraints for gauge interaction associated with $Z'$.
The gauge coupling and $Z'$ mass are restricted by the neutrino trident process $\nu N \to \nu N \mu^+ \mu^- $ where $N$ is a nucleon~\cite{Altmannshofer:2014pba}. 
The approximated bound is given by $m_{Z'}/g' \gtrsim 550 \ {\rm GeV}$ for $m_{Z'} > 1$ GeV, 
and we apply the bound in our numerical analysis below.

The gauge interaction is also constrained by the LHC experiment searching for the signal of $pp \to \mu^+ \mu^- Z' (\to \mu^+ \mu^-)$ as given in ref.~\cite{Sirunyan:2018nnz}.
The experimental results put the constraint on new gauge coupling in the mass range $5 \lesssim m_{Z'} \lesssim 70$ GeV.
We will compare parameter region explaining $B$ anomalies with the constraint.
}

\subsection{Decay of charged scalar \label{sec:Sdecay}}

Finally, we discuss the decay of charged scalar that provides implication to collider physics when we introduce $U(1)_{\mu -\tau}$ symmetry. 
As we discussed below Eq.~\eqref{Eq:lag-yukawa} charged scalar decays into only third generation of leptons when we chose $x=-2$.
We then chose $x=-2$ to relax collider constraint from charged scalar signature.
The decay width of $s^+$ for $x = -2$ is given by
\begin{equation}
\Gamma_{s^\pm \to \tau_R^\pm \nu_{R_\tau}} \simeq \frac{k_{33}^2}{16 \pi} m_{S},
\end{equation}
where we ignored lepton mass in the final state assuming light right-handed neutrino.
Also we assume right-handed neutrinos are long-lived and it will be just missing energy at collider experiments.
Note also that the lightest particle among $Q'$, $L'$ and $s^+$ would be stable when there is no interaction associated with $h_{ij}$ or $k_{ij}$ in other choices of $x$ value.

{
\section{Numerical analysis}
\begin{figure}[t]
\centering
\includegraphics[width=90mm]{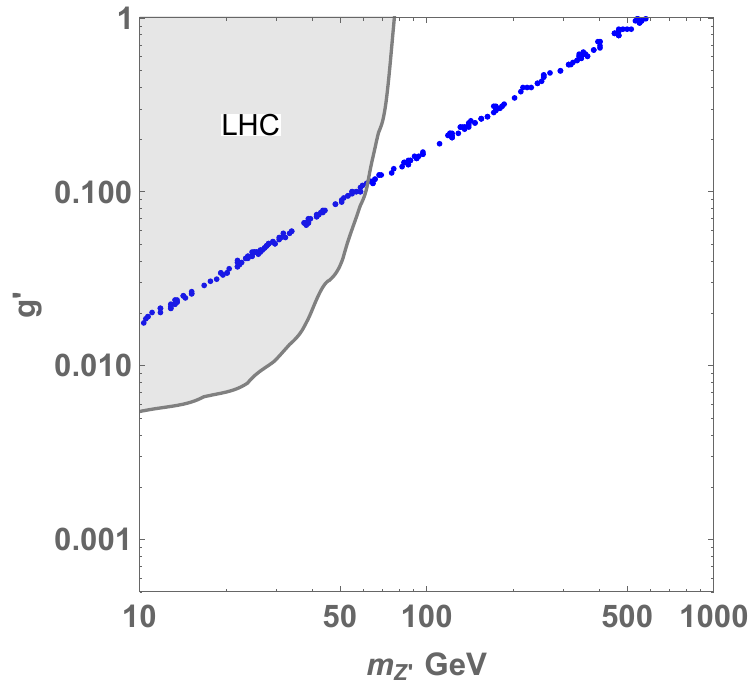} 
\caption{
Allowed points in the parameter space of $m_{Z'}$ and $g'$ which can explain the anomaly of $b\to s\mu\bar\mu$, providing  
 $\Delta C_9$ within 1$\sigma$ interval of global fit.
We also show the region excluded by $pp \to \mu^+ \mu^- Z' (\to \mu^+ \mu^-)$ search at the LHC experiment.}
\label{fig:C9Zp1}
\end{figure}
In this section we perform a numerical analysis to search for parameter sets which accommodate all the phenomena discussed above.
Here we scan our relevant free parameters $\{g_{ia}, f_{2a}, g', m_{Z'}, M_{\psi_a}, M_{Q_a}, m_S\}$ globally in the following range:
\begin{align}
& g_{ia} \in [10^{-5}, 1], \ f_{2a} \in [10^{-2}, 1], \ g' \in [10^{-3}, 1], \ m_{Z'} \in [10, 1000] \ {\rm GeV}, \nonumber \\
&  M_{\psi_1} \in [100, 500] \ {\rm GeV}, \ M_{\psi_2} \in [M_{\psi_1}, 750] \ {\rm GeV}, \ M_{\psi_3} \in [M_{\psi_2}, 1000] \ {\rm GeV} \nonumber \\
&  M_{Q'_1} \in [1000, 5000] \ {\rm GeV}, \ M_{Q'_2} \in [M_{Q'_1}, 5000] \ {\rm GeV}, \ M_{Q'_3} \in [M_{Q'_2}, 5000] \ {\rm GeV} \nonumber \\
& m_{S} \in [M_{\psi_1}-20, M_{\psi_1} -10] \ {\rm GeV},
\end{align} 
where we also chose $x = -2$ for $U(1)_{\mu - \tau}$ charge assignment.
Here we chose values of $M_{\psi_1}$ and $m_S$ to be {nearly} degenerated to avoid constraints from heavy charged lepton search at collider experiments.
We find that $b\to s\mu\bar\mu$ and the neutral meson mixing mainly depends on following Yukawa coupling combinations:
\begin{align}
C_9^{\mu}\sim g_{21}g^*_{13} |f_{21}|^2,\
\Delta m_K \sim g_{21}g^*_{11},\
\Delta m_{B_s} \sim g_{31}g^*_{12},\
\Delta m_{B_d} \sim g_{31}g^*_{11}, \
\Delta m_{D} \sim g_{11}g^*_{12}.
\end{align}
Since we would like to increase $C_9^{\mu}$ as large as possible, while all the meson mixings 
{should be within the experimental ranges,}
the following hierarchy is preferred
\begin{align}
g_{11}<<g_{21}\lesssim g_{31}.
\end{align}
Then, we estimate $C_9^\mu$ and muon $g-2$ imposing experimental constraints.
In Fig.~\ref{fig:C9Zp1} we show allowed parameter space in terms of $m_{Z'}$ and $g'$ to explain the $b\to s\mu\bar\mu$ anomalies $via$ $\Delta C_9^{Z'}$ within $1\sigma$ region of global fit.
We also show the parameter region excluded by the LHC measurement searching for $pp \to \mu \bar \mu Z'(\to \mu \bar \mu)$ process~\cite{Sirunyan:2018nnz}.
We find that parameter region of $m_{Z'} \lesssim 50$ GeV is excluded by the LHC constraints while heavier $Z'$ region can accommodate the $B$ anomalies.
For allowed region, the upper limit of $g'$ for fixed $m_{Z'}$ is determined by constraint from neutrino trident while the lower limit is given by constraint from $BR(B^0_s \to \mu^+ \mu^-)$. 
As a result, we find narrow range of parameter space where region close to neutrino trident limit $m_{Z'}/g'>550$ is allowed.
Note that the maximum $|C_9^{\mu [{\rm box}]}|$ is 0.115 at most that is out of the 3$\sigma$ range of experimental result due to
the stringent constraint arising from $\Delta m_{B_s}$, because they ($\Delta C_9^{\mu [{\rm box}]}$ and $\Delta m_{B_s}$) are proportional to the same combination $g_{31}^{}g_{21}^{}$.~\footnote{If one extends $g_{ai}$ to be complex, then one can evade the constraint of $\Delta m_{B_s}$ and keep large value of $|\Delta C^\mu_9|$.
However, in this case, another experimental bound of CP asymmetry $A_{CP}$ arises and it gives more stringent constraint~\cite{DiLuzio:2019jyq}.} 
We thus need contribution from $Z'$ interaction to explain the $B$ anomalies.

Next we show muon $g-2$ for allowed parameter sets satisfying all experimental constraints and explaining the $B$ anomalies.
In left(center) plot of Fig.~\ref{fig:mg2-I}, we show contribution to muon $g-2$ from scalar($Z'$) loop as a function of $M_{\psi_1}(m_{Z'})$, and 
total muon $g-2$ is shown in the right plot of the figure.
We find that contribution from $Z'$ loop can be larger than $2 \times 10^{-10}$ for $m_{Z'} \lesssim 600$ GeV.
{Notice here that the upper bound up to $600$ GeV comes from $m_{Z'}/g'>550$ GeV while that above $600$ GeV comes from our choice of $g' < 1$. }
The contribution from scalar loop can be larger than $10^{-10}$ for $m_{\psi_1} \lesssim 260$ GeV.
In particular it can be close to $10^{-9}$ for $m_{\psi_1} \sim 100$ GeV region.
It is thus possible to explain muon $g-2$ within $2 \sigma$ level when we add both $Z'$ and scalar contributions for light $m_{\psi_1}$ region.
\begin{figure}[tb]
\begin{center}
\includegraphics[width=5.35cm]{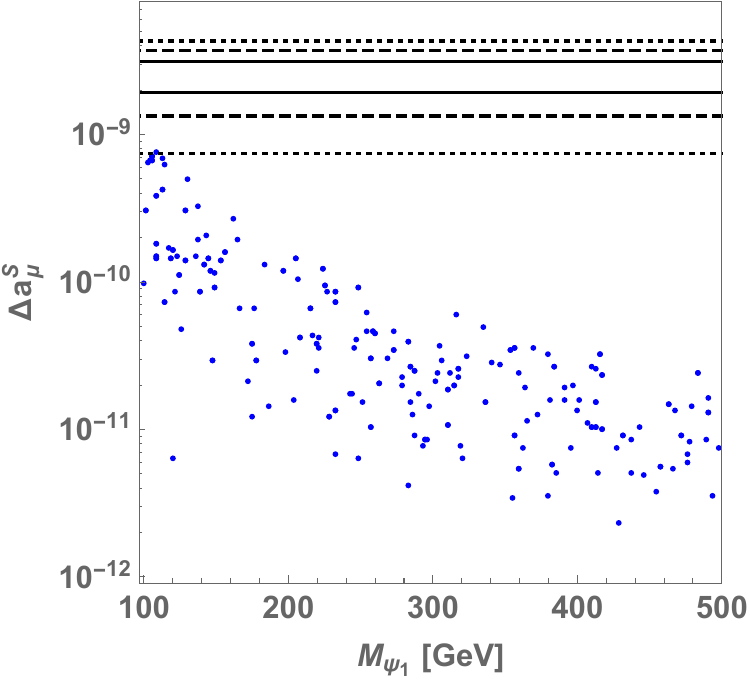}
\includegraphics[width=5.35cm]{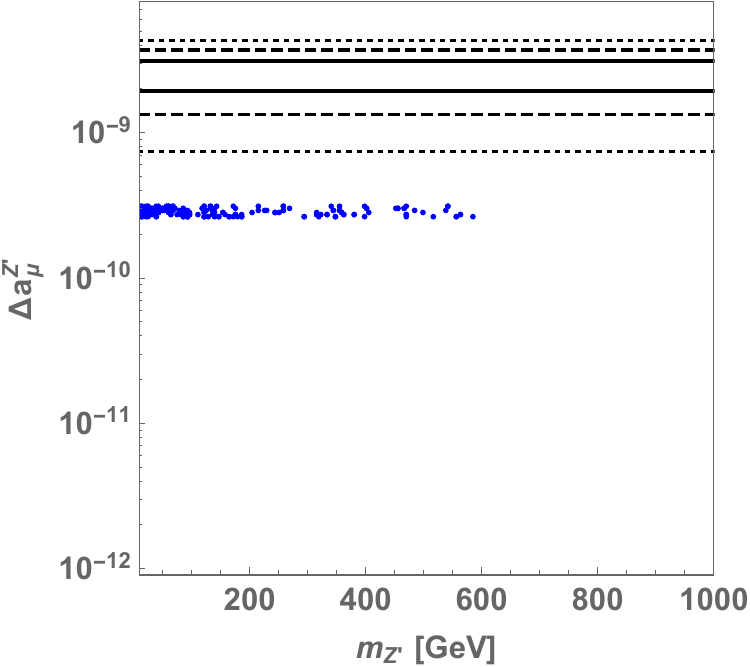}
\includegraphics[width=5.35cm]{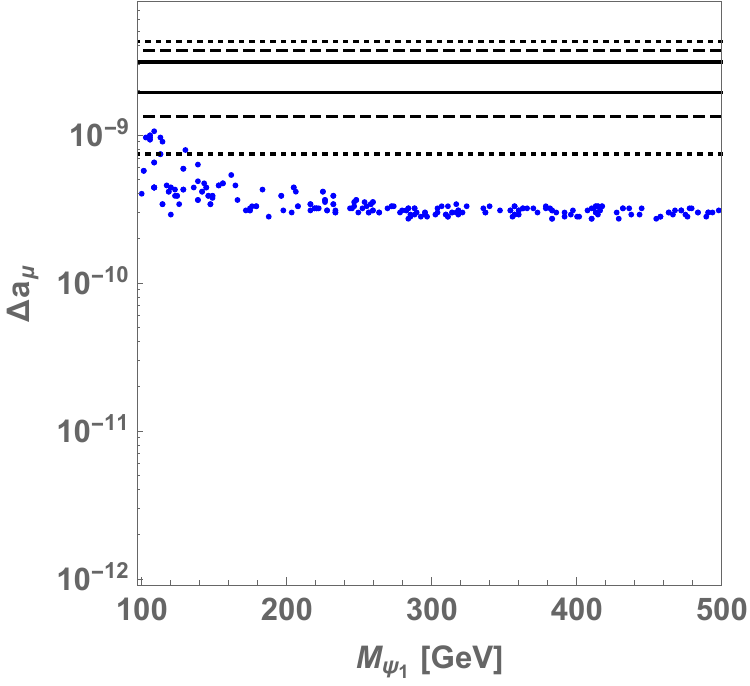}
\caption{Left : Contribution to muon $g-2$ from scalar loop diagrams. Center : Contribution to muon $g-2$ from the $Z'$ loop diagram. Right: Sum of scalar and $Z'$ contributions. The regions between solid, dashed and dotted lines indicate 1 $\sigma$, 2 $\sigma$ and 3 $\sigma$ region of deviation between observed value and the SM prediction respectively.} 
\label{fig:mg2-I}
\end{center}
\end{figure}
We also note that the upper bound on $f_{21}$ is $\sim 0.6$ which restrict maximum value of $\Delta a_\mu^{S}$.
{Here this upper bound of $f_{21}$ originates from the constraints of $Z$ boson decays.}

}


\subsection{Collider physics and constraints}
As discussed in the previous subsections, in order to get sizable muon $g-2$ satisfying the 
flavor constraints together, 
the mass scale (M) of exotic lepton doublet is required to be light; 
to obtain $\Delta a_\mu^S \gtrsim \mathcal{O}(10^{-10})$ we need $M \lesssim 300$ GeV.
 Here we are interested in the production and decay modes of the 
 doubly charged vector like lepton (VLL) given by,\\ 
\begin{center}
$p p \rightarrow \psi^{++}\psi^{--}, \psi^{++} \rightarrow (\mu ^{+} s^{+})\rightarrow \mu ^{+}(\nu_l l^{+})$, \\
~~~~~~~~~~~~~~~~~~~$\psi^{--} \rightarrow (\mu ^{-} s^{-})\rightarrow \mu ^{-}(\bar \nu_l l^{-})$.\\
\end{center}
Hence the final state is 1 oppositely charged muon pair + 1 oppositely charged lepton ($l$) + MET. 
As we choose $f_{21}=0.5$, $\psi^{\pm \pm}$ will decay mostly in to muon and a charged scalar. 
 Now the coupling of the charged scalar with the SM 
lepton and the neutrino is defined by $k_{33}$ as discussed in Sec.~\ref{sec:Sdecay}, 
and we consider it to be of the order 0.01 where $s^+$ decays into $\tau^+ \bar \nu_{R_\tau}$ with 100 $\%$ branching ratio.

Vector like leptons and quarks are constrained from the collider physics experiments. 
The ATLAS Collaboration performed a search for heavy 
lepton resonances decaying into a Z boson and a lepton in a multi lepton 
final state at a center-of-mass energy of 8 TeV~\cite{Aad:2015dha}, 
constraining singlet VLL model and excluding its mass range of 114$-$176 GeV. 
For the doublet VLL model, the L3 Collaboration at LEP placed a 
lower bound of $\sim$ 100 GeV on additional heavy leptons~\cite{Achard:2001qw}. 
Ref.~\cite{Sirunyan:2019ofn} and \cite{CMS:2018cgi} have shown that 
the VLL's in the mass range 120 $-$ 740 GeV are excluded with 95\% CL in 
different multilepton signals. In those analysis, the vectorlike 
leptons were singly charged and hence it only decays to a SM boson ($H$, $W$, $Z$) and SM 
leptons. While, in our case, VLL's decay in to a charged scalar and muon 
specifically, followed by the decay of the off-shell or on-shell 
charged scalar into neutrino and another $\tau$ lepton. 
Here we assume $M_{\psi} \sim m_{S}$ and produced muon is less energetic which would be missed at detectors by kinematical cut.
Hence the characteristic of 
our signal is significantly different from Ref.~\cite{Sirunyan:2019ofn} 
and \cite{CMS:2018cgi}. Similarly, for vectorlike quarks, the current limit is 1-1.3 TeV \cite{Buckley:2020wzk}, 
but in our model it decays $via$ the charged scalar, hence resulting in different 
final states, not searched so far at LHC.

LEP experiment excludes the charged Higgs masses below 80 GeV~\cite{ALEPH:2013htx}. At the LHC, 
searches for the charged Higgs have been performed through various decay channels, 
$H^{\pm}\rightarrow cs$~\cite{ATLAS:2013uxj}, $tb$~\cite{ATLAS:2018ntn} and 
$\nu \tau^{\pm}$~\cite{CMS:2019bfg}, 
and most of these searches exclude $m_H^{\pm} < m_t$. Other searches such as \cite{CMS:2019bfg}
give upper limit on the cross section$\times$ BR as a function of the charged scalar mass. 
Notice that $s^\pm$ only pair produced via $Z/\gamma$ propagator in $s-$channel and the cross section is below the current limit.

In this analysis we choose our selections differently than Ref.~\cite{Sirunyan:2019ofn} 
and \cite{CMS:2018cgi}. As a small mass difference between the 
charged Higgs and the VLL is naturally implied from the muon ($g-2$), the 
muon will have a very small $pT(\sim 10$) GeV, but the other two leptons 
will have a much higher $pT$. 
Other two leptons are $\tau$ in our case since we chose $U(1)_{\mu -\tau}$ charge so that charged scalar couples only $\tau$ and $\tau$-neutrino. 
This scenario is still allowed for VLL mass $\leq 300 $ GeV. 
There are scenarios \cite{Ma:2014zda, Yu:2015pwa} when the 
doubly charged VLLs decays to a $W^{\pm}$ and 
and lepton($l^{\pm}$), giving a final state of 2 oppositely charges lepton 
pair ($l^{\pm}$) + MET. In this study we have focused on 
a more exotic scenario, as proposed by the $U(1)_{\mu - \tau}$ extended model, 
where the  charged exotic leptons decays to tau lepton and a 
neutrino $via$ the charged scalar.
Hence in this study we select our signal to be 1 oppositely charged muon 
pair with very small pT + 1 oppositely charged tau pair with moderate pT 
+ MET, and we keep the mass difference between the charged Higgs and 
the VLL $\sim$10 GeV. The same final state has also been studied 
for a more general model of vector like leptons in Ref.~\cite{Kumar:2015tna}. 
One of the advantages of 
VLL with small mass is that the cross section is large which can 
negate the effect of the suppression due to more than one tau tagging. 
Moreover, in the VLL signatures studied so far by CMS and ATLAS the assumption 
was that VLL decays to a W or Z, which is unlikely in our case. 
As a result, $W/Z$ veto can increase the signal efficiency. 

We write the model Lagrangian of Eq.~(\ref{Eq:lag-yukawa}) in FeynRules (v2.3.13) 
\cite{Alloul:2013bka,Christensen:2008py}. We generate the 
model file for MadGraph5\_aMC@NLO (v2.2.1)~\cite{Alwall:2014hca} using FeynRules. Then we 
calculate the production cross section using the NNPDF23LO1 parton 
distributions~\cite{Ball:2012cx} with the factorization and renormalization scales at the 
central $m_T^2$ scale after $k_T$-clustering of the event.
We have computed the signal cross section of { $p p \to \psi^{++} \psi^{--}$}, 
where $p = q, \bar q, \gamma$. The cross sections are normalised to the 5 flavor scheme.
The inclusion of the photon PDF 
increases the signal cross section significantly as the coupling 
is proportional to the charge of the fermion.
We plot the the production cross section in Fig.~\ref{fig:cross1} for 13 TeV as well as 27 TeV~\footnote{Production cross section $pp \to s^+ s^-$ is much smaller than that of $\psi^{++} \psi^{--}$ and mass region $M(m_S) < 150$ GeV 
is still allowed by current experimental constraint~\cite{CMS:2019bfg}.}.
\begin{figure}[tb]
\begin{center}
\includegraphics[width=9.0cm,height=7.0cm]{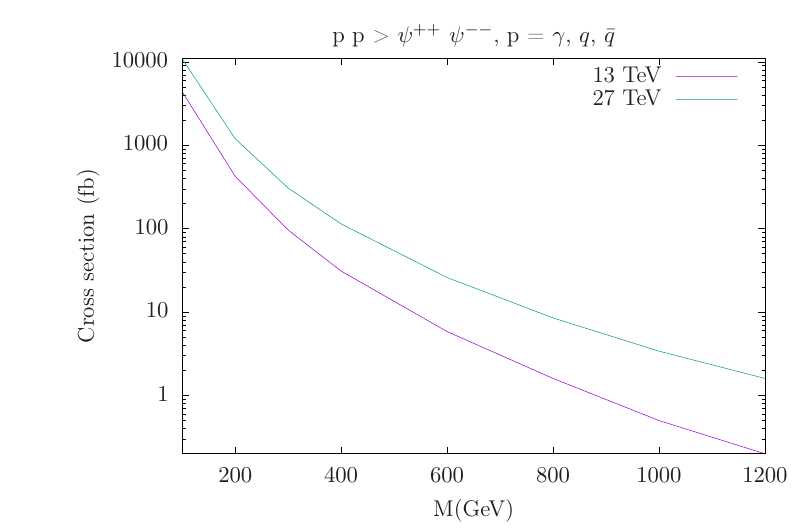}
\caption{The cross section for pair production process 
$p p \to \psi^{++} \psi^{--}$ as a function of VLL mass at 13 TeV and 27 TeV.}
\label{fig:cross1}
\end{center}
\end{figure}
After showering events in PYTHIA \cite{Sjostrand:2006za}, events are
passed through DELPHES~3 \cite{deFavereau:2013fsa} for detector 
simulation. In DELPHES, we choose the isolation cut for leptons to 
be $\Delta R_{max} = 0.5$, to ensure no hadronic activity inside 
this isolation cone. 
While generating the events, we kept the min $pT$ for muons to be 6 GeV, 
and also follow other trigger requirements for the soft muons 
following \cite{Sirunyan:2018iwl}. The tau tagging efficiency is considered to be 0.6 
and the misidentification efficiency is 0.01.

\begin{figure}[tb]
\begin{center}
\includegraphics[width=6.9cm,height=5.7cm]{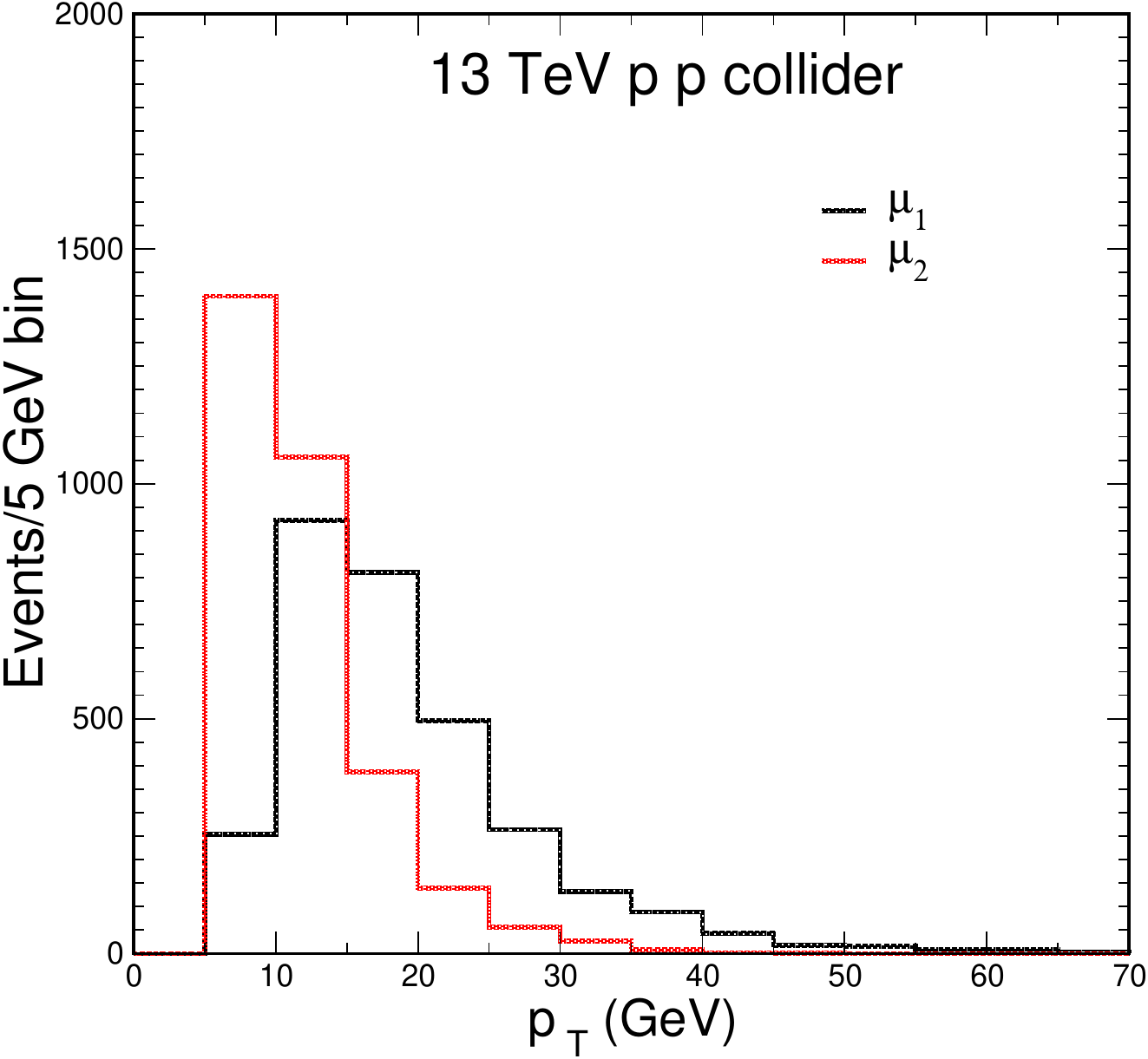}
\includegraphics[width=6.9cm,height=5.7cm]{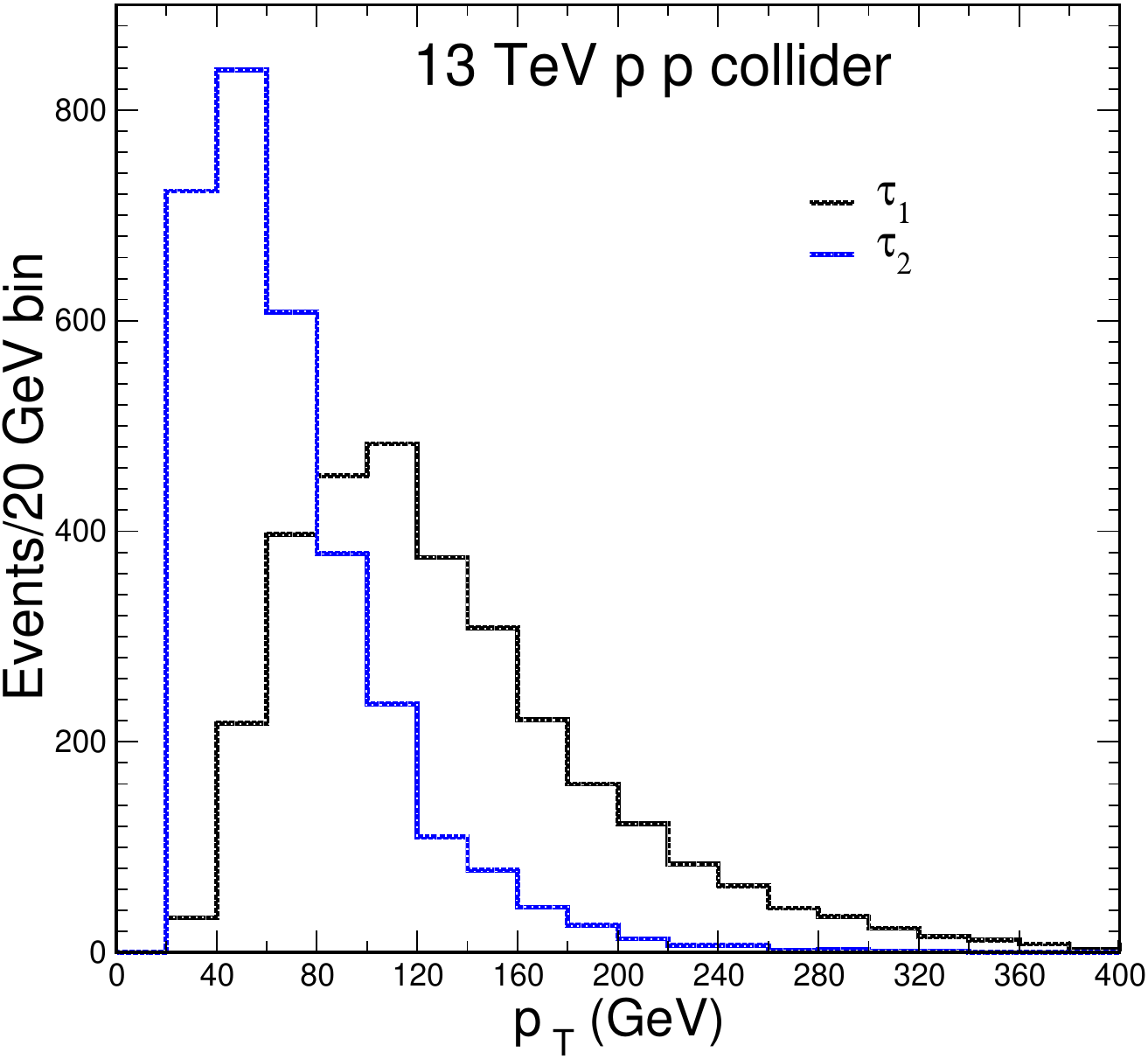}
\caption{The transverse momentum distribution of the leading(1) and the subleading(2) 
muon (left) and tau pairs (right) in unweighted events of {$p p\to \Psi^{++} \Psi^{--}$} at 13 TeV 
p-p collision for BP1.}
\label{fig:pT}
\end{center}
\end{figure}
\begin{figure}[tb]
\begin{center}
\includegraphics[width=6.9cm,height=5.7cm]{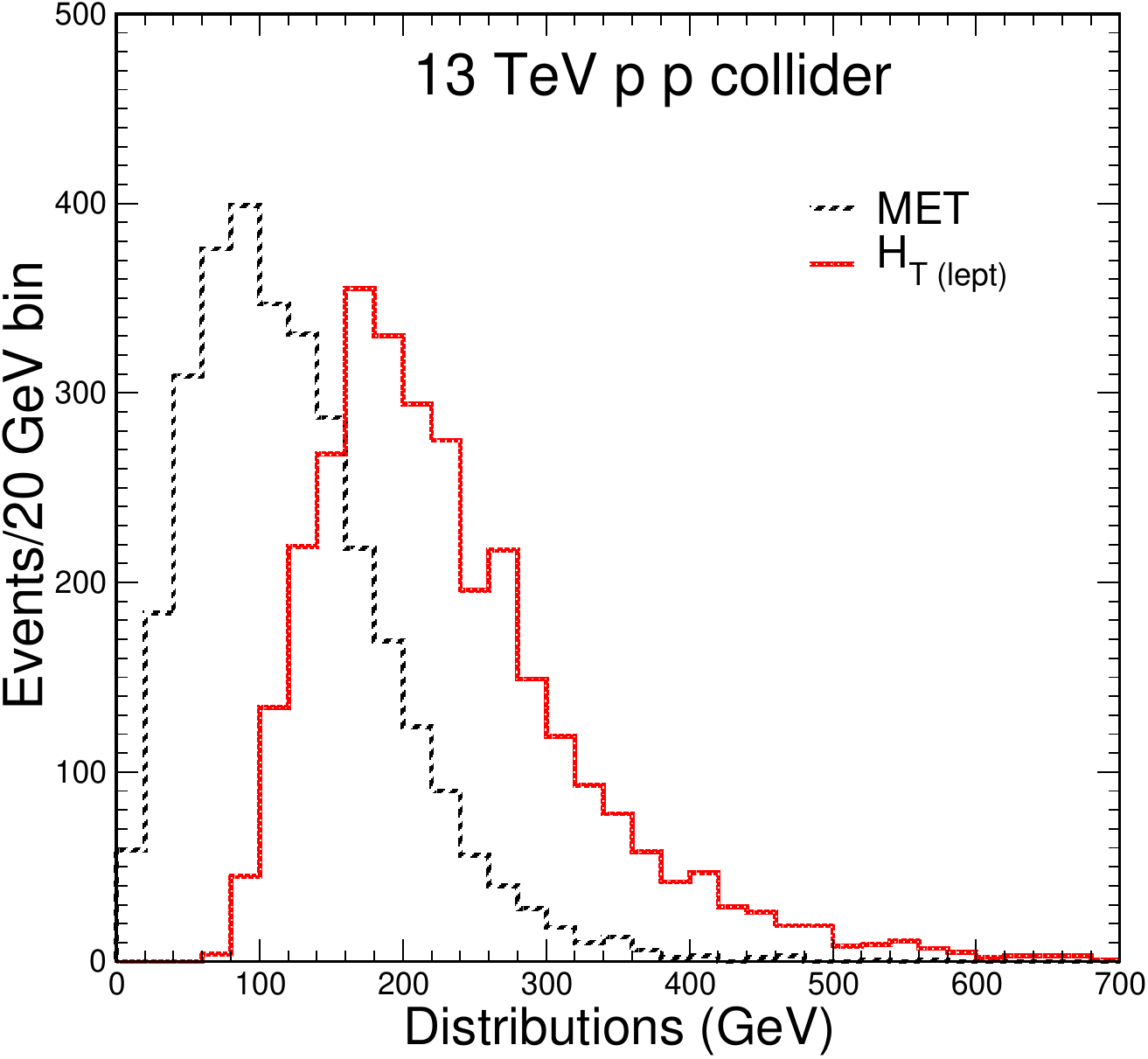}
\includegraphics[width=6.9cm,height=5.7cm]{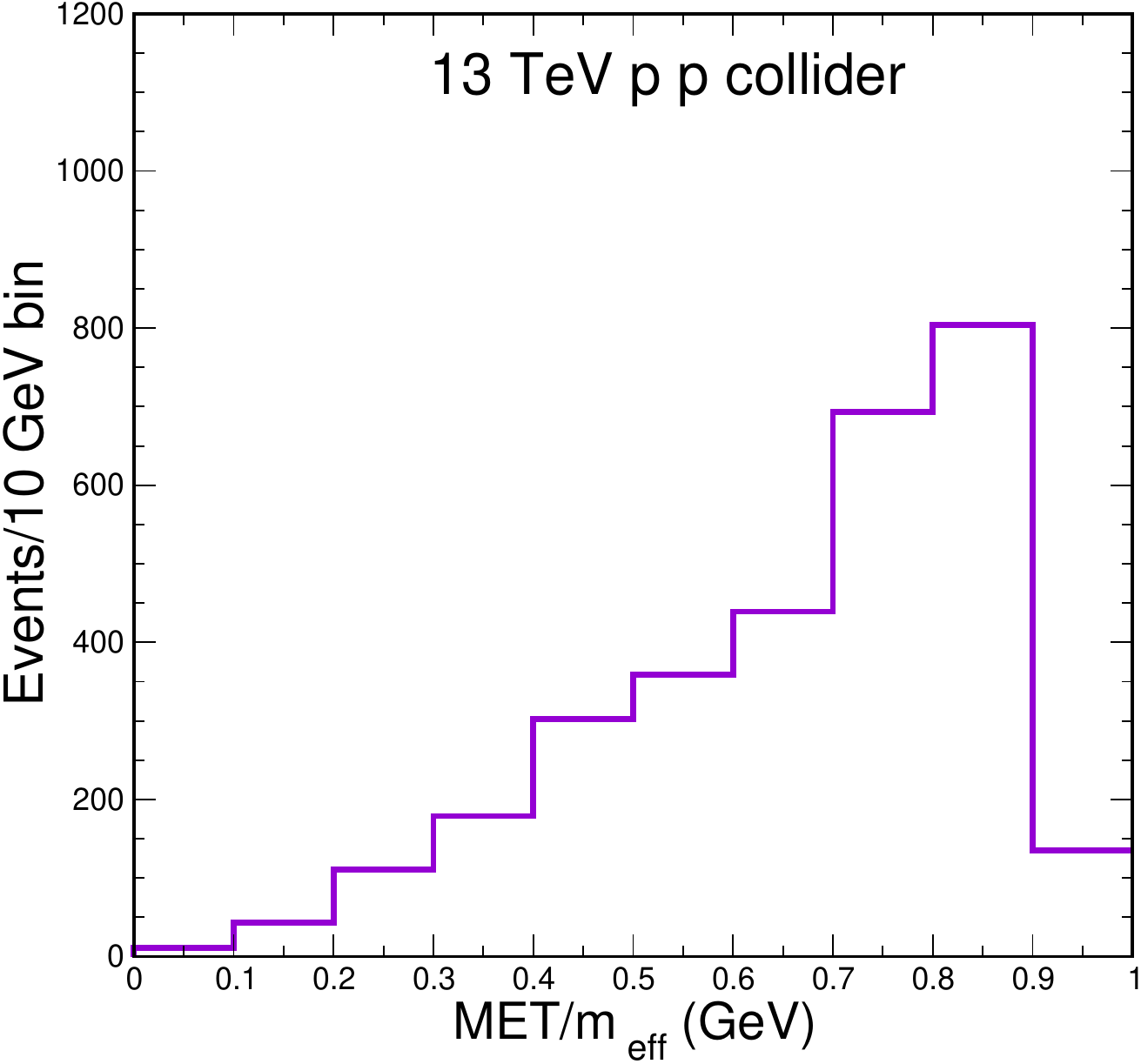}
\caption{The transverse missing energy (MET) and sum of all lepton $pT$ distribution is 
shown in the left. In the right the distribution is for the ratio of the MET and $m_{eff}$.
Events are unweighted and generated by {$p p \to \Psi^{++} \Psi^{--}$} at 
13 TeV p-p collision for BP1.}
\label{fig:MET}
\end{center}
\end{figure}
The $pT$ distribution of the leading and subleading tau and muon 
is shown in Fig.~\ref{fig:pT} for BP1. In Fig.~\ref{fig:MET} (left) we show the transverse 
missing energy and $H_T(l)= \sum_{i} p_T(l)_i$ distribution and (right)
the ratio $MET/m_{eff}$ ($m_{eff}= E_T+ H_T(l)+ H_T(j)$), which is effective to reduce the QCD-jet backgrounds. 
Based on these distributions 
we select a set of simple cuts on different kinematic variables. \\\\
{\bf Selection 1:}
\begin{itemize}
\item Opposite sign same flavor pair of mu and tau ($\mu^+\mu^-$) + ($\tau^+ \tau^-$),
\item $p_T(\mu_1)>$ 6 GeV, $p_T(\mu_2)>$6 GeV, $p_T(\tau_1)>$ 60 GeV, $p_T(\tau_2)>$40 GeV,
\item $|\eta (\mu,\tau)|< 2.5$, $\Delta R(l,l) > 0.3$,
\end{itemize}
{\bf Selection2:}
\begin{itemize}
\item b-jet veto, $MET >100$ GeV, $H_T>150$ GeV,
\item $MET/m_{eff} >0.5$,
\end{itemize}
{\bf Selection3:}
\begin{itemize}
\item Z veto with $M_Z\pm 10$ GeV.
\end{itemize}
We show the signal cross section after the selections in Table.~\ref{tab:3} for three BP's. One can see that 
for this multilepton channel the cross section is well above 1 fb after the selections. The 
signal does not suffer much from the Z-veto which is a big advantage for our signal as Z 
veto is effective to reduce the backgrounds from Z decays. b -jet veto and the requirement of 
higher ratio of MET and $m_{eff}$ will also be effective to reduce the background for these type 
 of signal. For the discussion on the background of this particular channel one can see 
Ref.~\cite{Kumar:2015tna}. In general multilepton channel possesses less background compared 
to the other processes. After Selection3, the number of events at 150 fb$^{-1}$
is always more than 150 if background is very small, which makes this channel a good candidate look for new physics at 13 TeV LHC run.
\begin{table}
\centering
\begin{tabular}{|c|c|c|c|}
\hline\hline
-&BP1 & BP2& BP3\\
\hline
$k_{33}=0.01$ &$g_{31}\approx-0.368$ ,$M_{Q'} =1083$   &$g_{31}\approx 0.32$ ,$M_{Q'} =1200$      &$g_{31}\approx-0.1080$ ,$M_{Q'} =1201$\\
             
&$g_{21} \approx 0.166$,$m_s \approx 272$ &$g_{21} \approx 0.2060$,$m_s \approx 230 $&$g_{21} \approx -0.6240$,$m_s \approx 304$\\
 
&$g_{11} \approx -0.0468$,$M \approx 284 $ &$g_{11} \approx -0.0014$,$M \approx 250$ &$g_{11} \approx 0.0071$,$M \approx 320$ \\

\hline\hline
Selection 1&3.44 (\it {9.58}) fb& 2.87 (\it {11.06}) fb& 2.67 (\it {9.62})fb\\
\hline
Selection 2 &1.76 (\it {7.31})fb &1.22 (\it {4.88})fb  & 1.49 (\it {4.36})fb\\
\hline
Selection 3 &1.63 (\it {5.82})fb &1.06 (\it {3.28})fb  & 1.38 (\it {4.96})fb\\
\hline
\end{tabular}
\caption{Signal cross section (fb) after the selections at three different benchmark points at 13 TeV and 
27 TeV (italic). Masses are in GeV.}
\label{tab:3}
\end{table}

\section{Conclusion}
{
We have analyzed muon (g-2), LFVs, $Z$ decays, $\Delta C_9^\mu$ for $B$-anomalies, and $M$--$\bar M$ mixing in a framework of multi-charged particles which includes exotic scalars, leptons and quarks under the local $U(1)_{\mu - \tau}$.
Thanks to the gauge symmetry we can suppress the LFV process which could appear from Yukawa interactions among exotic lepton, charged scalar and the SM lepton.
As a result, we found that the sizable Yukawa couplings are naturally allowed to explain muon $g-2$. 
We have first formulated phenomenological observables mentioned above in our model and performed numerical analysis to search for allowed parameter sets.

Carrying out numerical calculations, 
we have found that our $\Delta C_9^\mu$ can accommodate $B$-anomalies where $Z'$ boson contribution is dominant.
On the other hand, contribution from box diagram in $\Delta C_9^{\mu [{\rm box}]}$ can only reach the value $\sim -0.1$ 
when we impose constraints from $Z\to\nu_i\bar\nu_j$ invisible decay, $Z\to\mu\bar\mu$ decay and $B_s$--$\bar B_s$ mixing.
This is due to the stringent constraints from $B_s$--$\bar B_s$ mixing and $Z\to\mu\bar\mu$ which restrict the relevant Yukawa coupling constants. 
We have shown that the muon $g-2$ in our model is sum of the contributions from scalar boson loop and $Z'$ loop diagrams.
It has been found that we can explain muon $g-2$ within $2 \sigma$ level when we include both these contributions.
Finally, we have studied the collider physics focusing on the production of doubly charged leptons using some benchmark points allowed by the numerical analysis. We have shown that the channel with pairs of oppositely charged muon and tau has some unique features that distinguish our model signatures from other vector like lepton signatures at LHC. The exotic vector like quarks and the $Z'$ will also give interesting collider phenomenology but we keep that for future study. 
}

\section*{Acknowledgments}
This research was supported by an appointment to the JRG Program at the APCTP through the Science and Technology Promotion Fund and Lottery Fund of the Korean Government. This was also supported by the Korean Local Governments - Gyeongsangbuk-do Province and Pohang City (H.O.). H. O. is sincerely grateful for the KIAS member, and log cabin at POSTECH to provide nice space to come up with this project. 
N.K. acknowledges 
the support from the Dr. D. S. Kothari Postdoctoral scheme (201819-PH/18-19/0013).
N. K. also acknowledges ``(9/27-28 @APCTP HQ) APCTP Mini-Workshop 
- Recent topics on dark matter, neutrino, and their related phenomenologies" where the 
problem was proposed and also thanks the hospitality of APCTP, Korea.



\end{document}